# ARTICLE INFORMATION

**Article title**

Dataset of polarimetric images of mechanically generated water surface waves coupled with surface elevation records by wave gauges linear array.

**Authors**

Noam Ginio[1], Michael Lindenbaum[2, 3], Barak Fishbain[1] and Dan Liberzon[1, 3].

**Affiliations**

[1] Faculty of Civil and Environmental Engineering, Technion, Haifa 320003, Israel
[2] Faculty of Computer Science, Technion, Haifa 320003, Israel
[3] Interdisciplinary Program for Marine Engineering, Technion 320003, Haifa, Israel

**Corresponding author's email address and Twitter handle**

E-mail: noamginio@campus.technion.ac.il

**Keywords**

Spatio-temporal measurements; water surface elevation; water waves; light polarization; polarimetric camera; light reflection.

**Abstract**

Effective spatio-temporal measurements of water surface elevation (water waves) in laboratory experiments are essential for scientific and engineering research. Existing techniques are often cumbersome, computationally heavy and generally suffer from limited wavenumber/frequency response. To address these challenges a novel method was developed, using polarization filter equipped camera as the main sensor and Machine Learning (ML) algorithms for data processing [1,2]. The developed method training and evaluation was based on in-house made supervised dataset. Here we present this supervised dataset of polarimetric images of the water surface coupled with the water surface elevation measurements made by a linear array of resistance-type wave gauges (WG). The water waves were mechanically generated in a laboratory waves basin, and the polarimetric images were captured under an artificial light source. Meticulous camera and WGs calibration and instruments synchronization supported high spatio-temporal resolution. The data set covers several wavefield conditions, from simple monochromatic wave trains of various steepness, to irregular wavefield of JONSWAP prescribed spectral shape and several wave breaking scenarios. The dataset contains measurements repeated in several camera positions relative to the wave field propagation direction.

# SPECIFICATIONS TABLE

| | |
|---|---|
| **Subject** | Applied Machine learning, Water Science and Technology, Computer Vision and Pattern Recognition. |
| **Specific subject area** | Water waves measurements, including polarimetric images of light reflection from the water surface and instantaneous water surface elevation obtained by wave gauges array. |
| **Type of data** | Raw wage gauges array measurements in Tables (.csv), Raw polarimetric images (.tiff), Raw checkerboard images (.tiff), ground point control points illustration (.jpg), WGs coordinates (.xlsx), summarizing table of dataset subsets (.xlsx). |
| **Data collection** | A total of 170,298 polarimetric images, in 54 sequences of artificial light reflections from water surface waves. The unidirectional propagating waves were mechanically generated in controlled laboratory experiments. Measurements of the instantaneous water surface elevation fluctuations were collected by resistance-type WGs at 128 Hz sampling frequency. The 2448 × 2048 pixels images were collected at 32 fps by a polarimetric camera equipped with linear polarization filters. Several camera positions and orientations relative to the wave propagation direction were covered. For each camera location, images of standard checkerboard were taken with measured ground control points to enable camera spatial calibration. |
| **Data source location** | T-SAIL laboratory, Faculty of Civil and Environmental Engineering, Technion, Haifa 320003, Israel. |
| **Data accessibility** | Repository name: ScienceDB<br><br>Data identification number: https://cstr.cn/31253.11.sciencedb.13968<br><br>Direct URL to data: https://doi.org/10.57760/sciencedb.13968 |
| **Related research article** | |

# VALUE OF THE DATA

- The dataset covers a wide range of surface waves characteristics and camera positions/orientations, contributing to the development of machine learning models for water waves measurements. By incorporating a wide range of wavefield characteristics, we increase data variety, contributing to creation of robust and reliable statistical models to learn the light polarization-to-slope transfer function.
- Due to the complexity of accurately measuring wavefields with directional energy propagation spread, our dataset provides several unidirectional subsets repetitions in various camera orientations relative to wavefield propagation direction, resulting in variety of polarimetric light reflection and wave slopes data.
- The use of in-house made artificial light source supported capturing polarimetric images at high spatial resolution and high signal to noise ratio (SNR).

# BACKGROUND

Water surface waves, driven by wind forcing, play a key role in marine processes, affecting energy transfer, sediment transport, and climate. Accurate wave monitoring is essential for environmental and engineering applications [3–5] and the ability to perform accurate measurements of complex wind forced waves is also required for research. However, existing methods are limited by low spatial resolution, frequency response, or high computational costs [6–8]. Photometric techniques, such as stereo imaging, also require extensive calibration and post-processing [9–12].

To overcome these limitations, we developed the Wave (from) Polarized Light Learning (WPLL) method, which uses deep neural networks to derive surface slopes from polarized light [1], and to integrate the slopes into surface elevation maps, effectively providing accurate spatio-temporal measurements of surface waves. The method resolves the full spectrum of wave frequencies/wavenumbers without relying on any simplifying assumptions is based on out proof-of-concept study [2]. Here we are publicly sharing the supervised dataset used for the development and evaluation of the WPLL method [13] as an applicable laboratory tool. This dataset contains high-resolution spatio-temporal water waves measurements coupled with polarimetric imaging of light reflection from the water surface. The dataset covers various wavefield and camera orientation conditions. This dataset will enable further modifications of the method in future research and can be used by other researchers and engineers for purposes of developing measurement techniques, testing relevant theories and more.

# DATA DESCRIPTION

The dataset consists of 54 subsets of wavefields experiments. Further description for each of these is provided in **Table *1*** below and in the "EXPERIMENTAL DESIGN, MATERIALS AND METHODS" section.

**Table 1.** Overview of 54 supervised datasets; camera locations; wavefield type and main parameters. MW refers to monochromatic wave train; SW refers to JONSWAP [14] prescribed power density spectral shape; BW refers to a wavetrain containing a wave breaking event. *Loc* refers to camera location and orientation relative to the wave propagation direction.

| # | WG file mark | Experiment Date | Loc | Wave field parameters | | | length |
|---|---|---|---|---|---|---|---|
| | | | | wave type | # frames | f / fp [Hz] | # frames |
| 1 | A | 14/06/2023 | $Loc_A$ | MW | 2.5 | 1 | 960 |
| 2 | B | | | | 3 | | 960 |
| 3 | D | | | | 2.7 | | 960 |
| 4 | E | | | | 2.2 | | 960 |
| 5 | G | | | | 2 | | 960 |
| 6 | H | | | | 1.7 | | 960 |
| 7 | J | | | | 1.5 | | 960 |
| 8 | K | | | | 1.2 | | 960 |
| 9 | L | | | | 1 | | 960 |
| 10 | M | | | | 0.7 | | 960 |
| 11 | N | | | | 0.5 | | 960 |
| 12 | O | | | | 0.3 | | 960 |
| 13 | D | 03/08/2023 | $Loc_A$ | BW | 2 | 1 | 1178 |
| 14 | E | | | | 3 | | 1280 |
| 15 | F | | | | 4 | | 1280 |
| 16 | L | | $Loc_B$ | | 2 | | 1280 |
| 17 | M | | | | 3 | | 1280 |
| 18 | N | | | | 4 | | 1280 |
| 19 | A | 20/12/2023 | $Loc_A$ | SW | 1 | 2 | 2400 |
| 20 | B | | | | 1.5 | | 2400 |
| 21 | C | | | | 2 | | 2400 |
| 22 | D | | | | 1.5 | 1.5 | 3200 |
| 23 | E | | | | 2 | | 3200 |
| 24 | F | | | | 2.5 | | 3200 |
| 25 | G | | | | 2 | 1 | 4800 |
| 26 | H | | | | 2.5 | | 4800 |
| 27 | I | | | | 3 | | 4800 |
| 28 | M | | | | 2.5 | 0.8 | 6400 |
| 29 | K | | | | 3 | | 6400 |
| 30 | L | | | | 3.5 | | 6400 |
| 31 | M | 20/12/2023 | $Loc_B$ | SW | 1 | 2 | 2400 |
| 32 | N | | | | 1.5 | | 2400 |
| 33 | O | | | | 2 | | 2400 |
| 34 | P | | | | 1.5 | 1.5 | 3200 |

| 35 | Q | | | | 2 | | 3200 |
|----|---|---|---|---|-----|-----|------|
| 36 | R | | | | 2.5 | | 3200 |
| 37 | S | | | | 2 | | 4800 |
| 38 | T | | | | 2.5 | 1 | 4800 |
| 39 | U | | | | 3 | | 4800 |
| 40 | V | | | | 2.5 | | 6400 |
| 41 | W | | | | 3 | 0.8 | 6400 |
| 42 | X | | | | 3.5 | | 6400 |
| 43 | A | | | | 1 | | 2400 |
| 44 | B | | | | 1.5 | 2 | 2400 |
| 45 | C | | | | 2 | | 2400 |
| 46 | D | | | | 1.5 | | 3200 |
| 47 | E | | | | 2 | 1.5 | 3200 |
| 48 | F | 28/12/2023 | $Loc_C$ | SW | 2.5 | | 3200 |
| 49 | G | | | | 2 | | 4800 |
| 50 | H | | | | 2.5 | 1 | 4800 |
| 51 | I | | | | 3 | | 4800 |
| 52 | M | | | | 2.5 | | 6400 |
| 53 | K | | | | 3 | 0.8 | 6400 |
| 54 | L | | | | 3.5 | | 6400 |

The entire dataset is openly accessible in the repository's data folder [13]. The structure of the data repository is detailed in **Figure 1** There are four subfolders, corresponding the data of each experiment. Next division is according to the camera location/orientation *Loc*. For each *Loc* there is calibration & WG data folder which contains: checkerboard images (.tiff), an illustration presenting the checkerboard location in WCS (.jpg), WGs locations (.xlsx) and WG time series data (.csv). Within each *Loc* subfolder the polarimetric images of each experiment are placed inside a dedicated folder. The entire images sequence of each experiment is compressed into a single archive file (.zip). A summary table of the 54 experiments (.xlsx) and the data folder layout presented in **Figure 1** (.jpg) are also stored in the main folder.

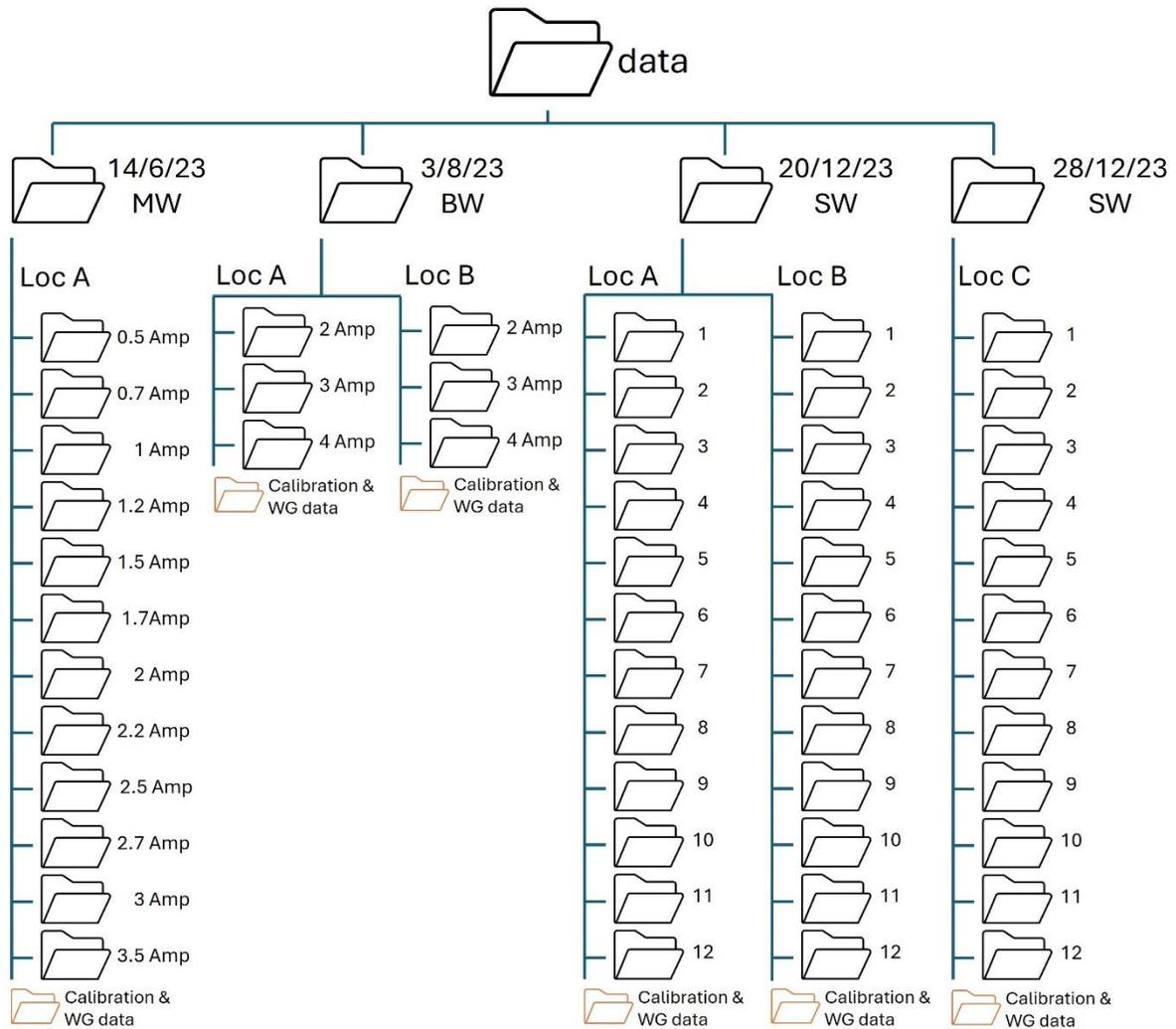

**Figure 1.** Data repository structure layout.

# EXPERIMENTAL DESIGN, MATERIALS AND METHODS

The experiments were conducted in the Technion Sea-Atmosphere Interactions Research Laboratory (T-SAIL) wave basin. The basin dimensions are $6 \times 6\ meters$ and the water depth, $h$, was kept at $70\ cm$. Flap-type mechanical wavemakers[1] (Edinburgh Designs®) were used to generate unidirectional waves. The basin was equipped with wave energy absorber opposite to the wavemakers, with the sidewalls left exposed (see **Figure 2**).

---

[1] Edinburgh Designs - Wave Generator *http://www.edesign.co.uk/product/ocean-flap-wave-generator/*

Two main types of propagating waves were generated: monochromatic wavetrains (MW) and irregular wavefields of prescribed JONSWAP [14] spectral shape, denoted as SW. The SW subsets served as an approximation of wind-forced waves. Additionally, sets of wavetrains containing breaking waves, referred to as BW, were generated by means of linear focusing. The BW subsets contained extreme steepness instances, constituting slope values outside the range of those in the MW and SW trains. Measurements of the water surface elevation fluctuations were collected by a linear array of eight resistance-type WGs, positioned at 12 *cm* intervals, as shown in **Figure 2**.

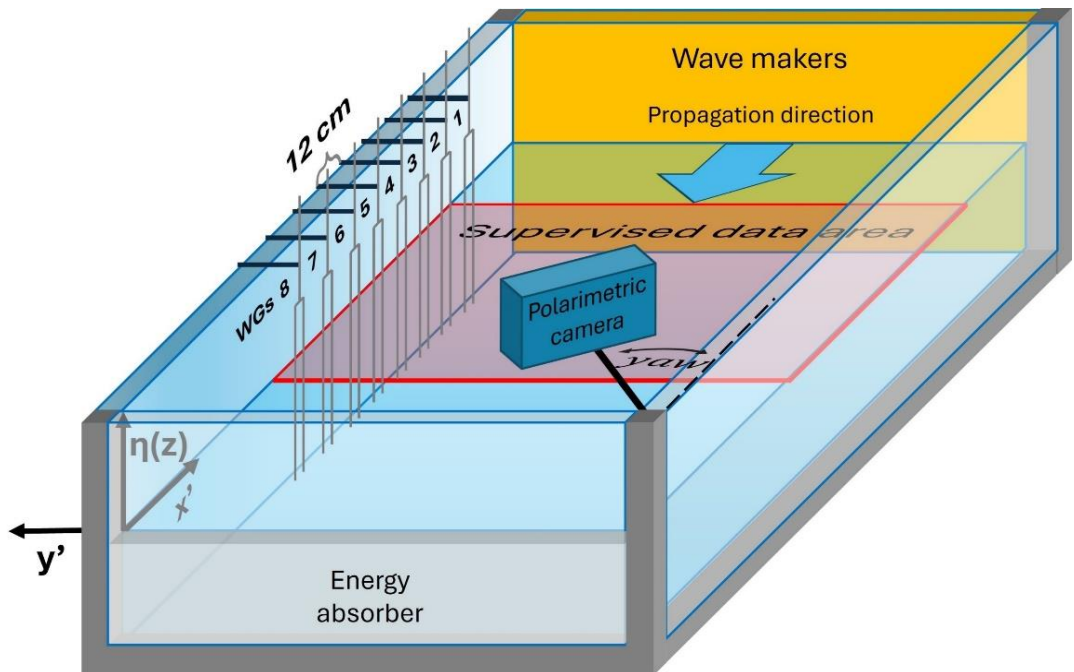

**Figure 2.** Experiment layouts of supervised data collection in the wave basin of polarized imagery depicting the spatial extent of the supervised data area bounded by the WG's locations.

The images were collected, at 32 *fps*, using a polarimetric camera (Sony IMX250MZR, 2448 × 2048 pixels, Polar-Mono model number BFS-U3-51S5M-C). A 12 *mm* focal length lens was used. Each pixel in this camera is equipped with a linear polarization filter (**Figure 3**(a)) fitted onto the camera sensor. The filters are arranged in periodic pattern, with each 2 × 2 pixel matrix containing filters oriented at *0°, 45°, 90°* and *135°*, as shown in **Figure 3**(b, c). The four recorded light intensities, after each polarization filter referred to as polarized light intensities and hereafter denoted as $I_0$, $I_{45}$, $I_{90}$, and $I_{135}$, respectively.

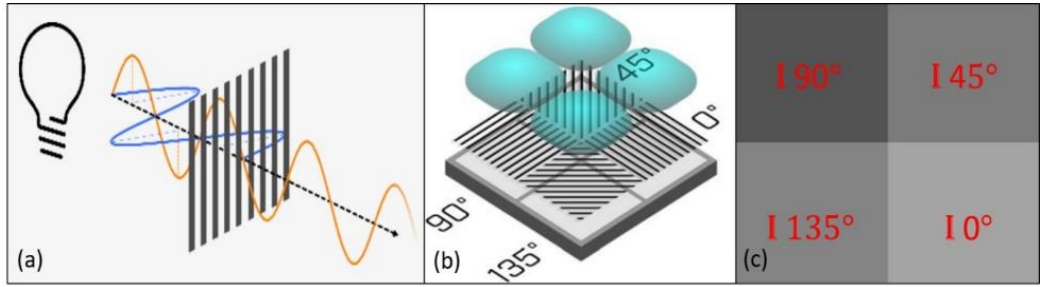

**Figure 3.** (a) The polarizing filter passes the beam that is aligned to the angle of the slits, and blocks the beam aligned perpendicular to them. (b) Each individual pixel has its own polarizing filter. (c) Zoom in image of the polarimetric image (Reproduced with permission from Teldyne FLIR).

A large artificial light source, hereafter denoted as $LArL$, was installed above the wave basin. The use of artificial light source allowed control over the reflected light intensity and masked out ambient light reflections, thus supporting high signal to noise ratio (SNR). The $2 \times 2$ meter$^2$ $LArL$ source was in-house made, utilizing 24 m long 12V flexible LED strip at 6000K color temperature (Inspired LED@) and was fitted with a light diffuser fabric (**Figure 4**).

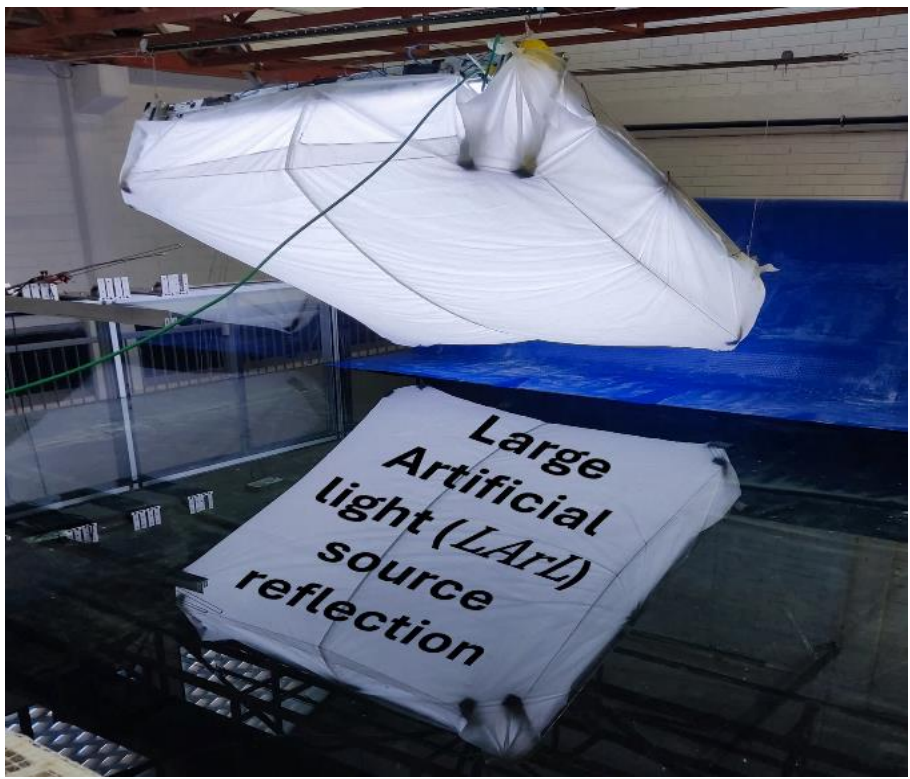

**Figure 4.** In-house made large artificial light (LArL) source reflection on the MWL. The black font notation is added in postprocessing to ease the reflection location identification by the reader.

The linear WGs array was positioned along the $x'$ axis, within the coordinates range of the $LArL$ reflection on the water surface (**Figure 2**). WGs provided records of the instantaneous water surface elevation fluctuations at $128\ Hz$, while the camera captured images at $32\ fps$. Synchronization between the image acquisition and the WGs was achieved using a signal generator to initiate measurements. To match the image acquisition rate for each experiment, the WGs data requires to be downsampled to correspond to a single frame's time span of 0.03125 seconds (32 Hz).

Three camera locations were selected on the basin wall, at significantly different positions along the negative $y'$ axis. At each location yaw, pitch, and elevation of the camera were set to maximize coverage of the reflected light patch at the water surface, keeping the linear WG array at the fringes of the imaged reflection patch, while the *LArL* remained stationary. The camera orientation at each location defined new orthogonal coordinate system on the water surface plane, with the $x'$ axis being the projection of the camera optical axis, and the origin location at the $y'$ axis of the corresponding Loc (See **Figure 5**). This rotated coordinate system, is denoted hereafter as RCS. Detailed parameters are listed in **Table 2**. The data was collected from three camera locations, denoted as $Loc_A$, $Loc_B$ and $Loc_C$. The MW trains, were collected in one location ($Loc_A$), the SW were collected in all three locations ($Loc_{A-C}$) and the BW sets were collected in two locations ($Loc_{A,B}$).

**Table 2.** Camera position and orientation parameters in the three experiment repetitions ($Loc$).

| Location (**Loc**) | $Loc_A$ | $Loc_B$ | $Loc_C$ |
|---|---|---|---|
| Yaw [deg] | $-15°$ | $-30°$ | $-39°$ |
| Pitch [deg] | $-42°$ | $-35°$ | $-36.3°$ |
| Roll [deg] | $4°$ | $4°$ | $4°$ |
| x' [cm] | 0 | 0 | 0 |
| y' [cm] | -257 | -360.5 | -469.5 |
| z' [cm] | 157 | 141.4 | 160.9 |

To use the polarimetric data from the image coordinate system (ICS) for inferring the water surface and slopes in the world coordinate system (WCS) a transformation between the image plane and the mean water level (MWL) is needed. To enable computation of this transformations, standard checkerboard images [15] and the checkerboard location in WCS coordinates (from which control points can be derived) are provided in each camera location. Additional projective transformations from the ICS to the various RCS-s corresponding to each camera location, $Loc_{A-C}$, are achievable by additional set of RCS control points coordinates in each location (see **Figure 5**). The three coordinate systems ICS, WCS and RCS-s use $(u, v)$, $(x', y')$ and $(x, y)$ coordinates, respectively.

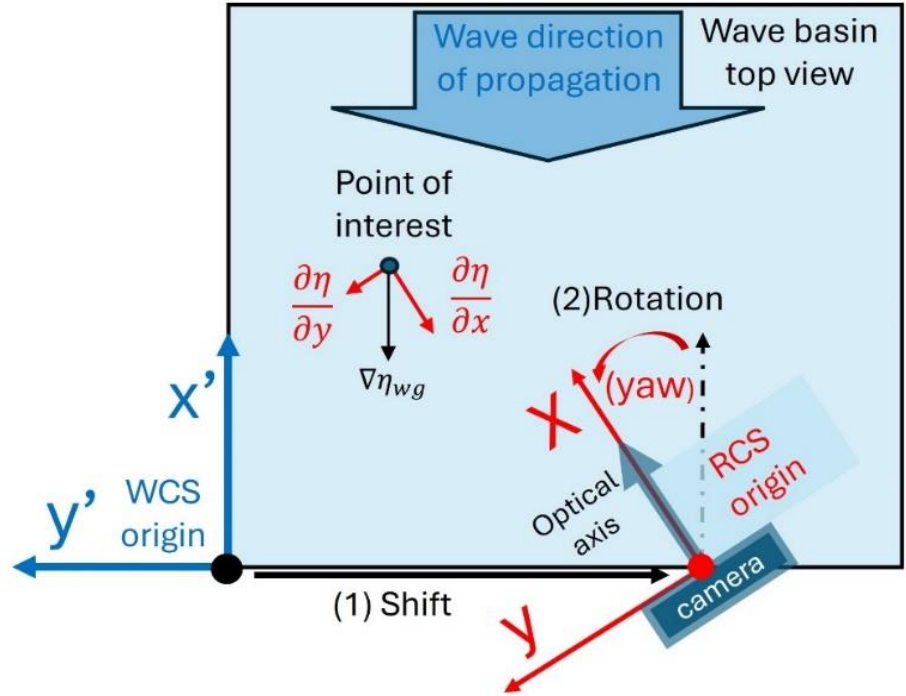

**Figure 5.** Illustration of ground truth data preperation by shift and rotation of WCS to RCS according to the camera location and orientation and vector decompostion of $\nabla\eta$.

The polarimetric camera, capturing the reflection of the polarized light from the water surface, produced 8-bit grayscale images. To ensure high SNR, the controllable light source was tuned so that the camera sensor's gamma factor remained at unity and the gain was set to zero. In addition, we optimized the camera aperture and exposure time, so the dynamic range is maximized, thus, minimizing cutoff and saturation.

Considering the experimental setup limitation, imposed mainly by the energy absorber efficiency, long-wavelength MWs, with a $1\ Hz$ frequency, were generated at various amplitudes to cover the range of slopes expected also in the SW fields. The 12 MW trains with amplitudes ranging from 0.3 cm to 3 cm in $0.2 - 0.3$ cm increments.

The 12 SW field subsets were created by generating irregular wave fields corresponding prescribed JONSWAP energy density spectral shape, formulated as

$$(1) \qquad S = \alpha_p \cdot g^2 \cdot 2\pi f^{-5} \cdot \exp\left[-\frac{5}{4} \cdot \left(\frac{f_p}{f}\right)^4\right] \cdot \gamma^{\exp\left(-\frac{(f-f_p)^2}{2\sigma^2 \cdot f_p^2}\right)}$$

where the parameters, describing the spectrum were: frequency $f = 1/T$ with T being the wave period, $\gamma$ being the peakiness coefficient and subscript p denoting the peak value, gravitational constant g, and spectral shape slopes $\sigma \in (0.0111$ for $f < f_P; 0.0143$ for $f \geq f_P)$ corresponding the long and the short waves ranges, $\alpha_p$ is the Phillip's equilibrium constant. The significant wave heights $H_s$ is then calculated by

$$(2) \qquad H_s = 4 \cdot \left(\int_0^\infty S(f)\, \partial f\right)^{0.5}.$$

Specifically, the parameters chosen for our experiments are given in **Table 3**.

**Table 3.** Parameters of the test set JONSWAP shaped spectrum waves (SW). the same parameters were used in each *Loc*.

| *#SW* | *peak frequency, $f_P$ [Hz]* | *peak time period, $T_p$ [sec]* | *peakiness coefficient $\gamma$* | *significant wave height, $H_s$ [cm]* | *dominant wavelength[2] [cm]* |
|---|---|---|---|---|---|
| *1* | | | | 2.5 | |
| *2* | 0.8 | 1.25 | 10 | 3 | 233.66 |
| *3* | | | | 3.5 | |
| *4* | | | | 2 | |
| *5* | 1 | 1 | 15 | 2.5 | 155.49 |
| *6* | | | | 3 | |
| *7* | | | | 1.5 | |
| *8* | 1.5 | 0.667 | 60 | 2 | 69.57 |
| *9* | | | | 2.5 | |
| *10* | | | | 1 | |
| *11* | 2 | 0.5 | 100 | 1.5 | 39.17 |
| *12* | | | | 2 | |

# LIMITATIONS

The WGs array provides time series of water surface elevation, allowing for interpolated curves representing surface elevation along the x' axis. The surface elevation can then be differentiated along the x' axis to generate local slope curves. Assuming that the mechanically generated waves are truly unidirectional, these curves reflect the wave and slope data. However, as the complexity of wavefields increases—particularly with shorter waves and sidewall reflections—energy is distributed in both the primary direction and laterally (y'). While polarimetric images capture these harmonics, WGs data struggle due to limited frequency/wavenumber response, constrained by WGs diameter and spacing. Moreover, the linear WGs array lacks information on wave components in the y' direction.

An example of this complexity is shown, in **Figure 6**, using images captured with a DSLR camera at a 40° yaw angle relative to the wave propagation (x') during $SW_{12}$. Spatial Fourier analysis of these images (2D FFT) [16], is presented in **Figure 7**. This analysis revealed spectral peaks along the x'-axis, indicating harmonics in the main propagation direction, and along the y'-axis, showing energy spread from standing waves. The red circle in marks the Nyquist frequency, underscoring the WG array's limitations in capturing higher harmonics.

---

[2] According to linear dispersion relation theory

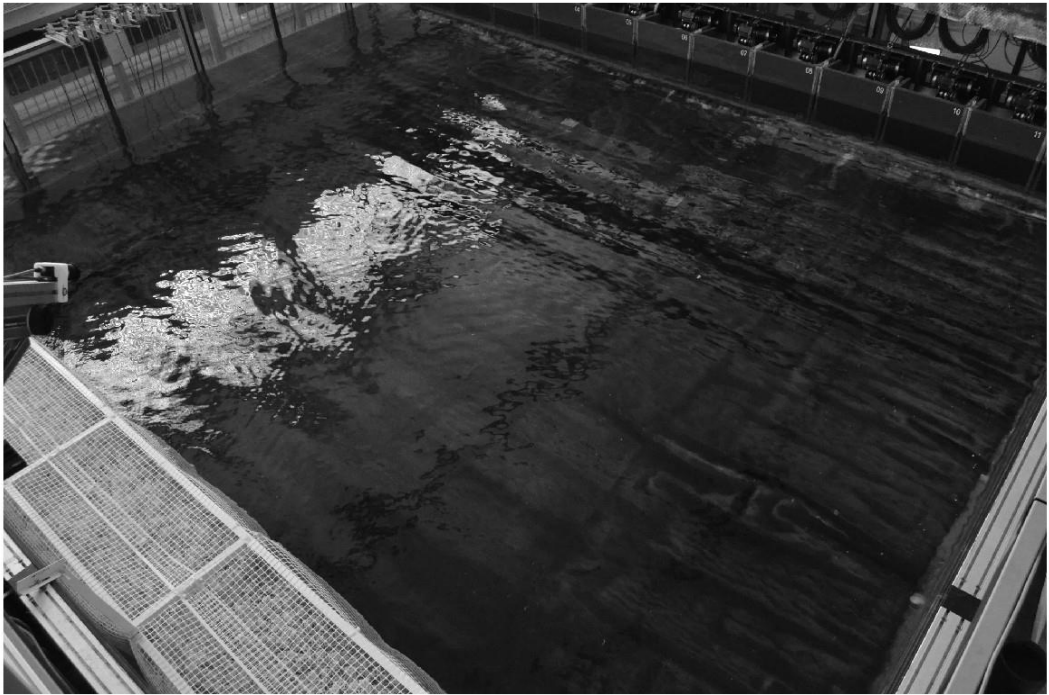

**Figure 6.** DSLR raw image covering the entire wave basin during experiment $SW_{12}$.

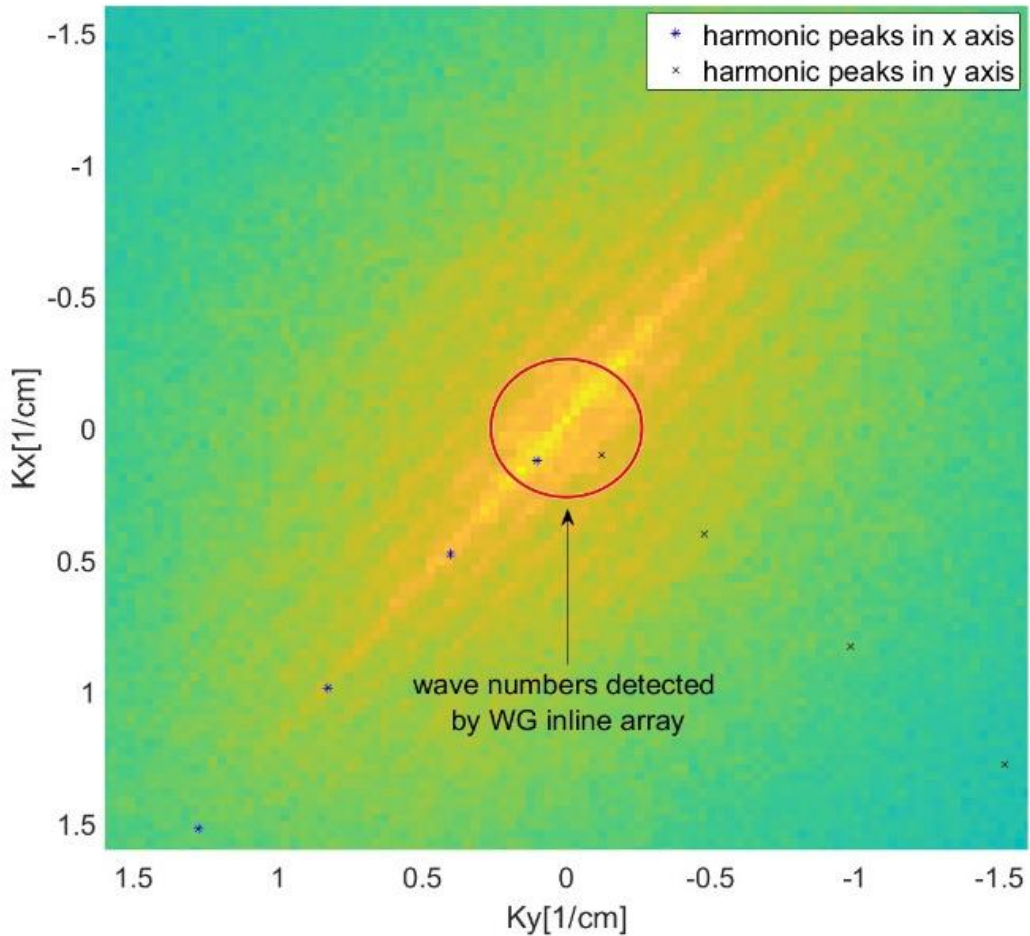

**Figure 7.** Mean light amplitude spectrum in the wavenumber domain calculated on the DSLR images in $SW_{12}$. The blue stars and black crosses represent the spectral peaks of the first four harmonics in $x'$ and $y'$ directions, respectively. The red circle represents a distance from the origin corresponding to Nyquist frequency for spacing between WG probes.

# ETHICS STATEMENT

Our study does not involve studies with animals or humans. We confirm that our research strictly adheres to the guidelines for authors provided by Data in Brief in terms of ethical considerations.

# CRediT AUTHOR STATEMENT

**Noam Ginio**: Conceptualization, Methodology, Software, Validation, Formal analysis, Investigation, Data curation, Writing – original draft, Writing – review & editing, Visualization; **Michael Lindenbaum**: Conceptualization, Supervision, Methodology, Validation, Writing – review & editing, Funding acquisition; **Barak Fishbain**: Conceptualization, Resources, Writing – review & editing, Supervision, Funding acquisition; **Dan Liberzon**: Conceptualization, Methodology, Validation, Resources, Writing – review & editing, Supervision, Project administration, Funding acquisition.

.

# ACKNOWLEDGEMENTS

The authors acknowledge the financial support provided by the Israeli Ministry of Energy (grant no.1016825) and the Israeli Ministry of Environmental Protection (grant no. 162-71). Noam Ginio is grateful for the research scholarships provided by the Israeli Port Company (IPC). A special thank goes to Ariel Weinstock and Jacob Mandel, for the aid and guidance in the construction of the large artificial light source that made this research possible.

.

# DECLARATION OF COMPETING INTERESTS

The authors declare that they have no known competing financial interests or personal relationships that could have appeared to influence the work reported in this paper.

# REFERENCES

[1]   Ginio, N., Lindenbaum, M., Fishbain, B., & Liberzon, D., 2024. Wave (from) Polarized Light Learning (WPLL) method: high resolution spatio-temporal measurements of water surface waves in laboratory setups. https://doi.org/10.48550/arXiv.2410.14988

[2]   Ginio, N., Liberzon, D., Lindenbaum, M., Fishbain, B., 2023. Efficient machine learning method for spatio-temporal water surface waves reconstruction from polarimetric images. Meas Sci Technol 34, **055801**. https://doi.org/10.1088/1361-6501/acb3eb

[3]   Tucker, M.J. and Pitt, E.G., 2001. *Waves in ocean engineering* (No. Volume 5).

[4]   Farmer, D.M., McNeil, C.L., Johnson, B.D., 1993. Evidence for the importance of bubbles in increasing air–sea gas flux. Nature 361, **620–623**. https://doi.org/10.1038/361620a0

[5]   Melville, W.K., Veron, F., White, C.J., 2002. The velocity field under breaking waves: coherent structures and turbulence. J Fluid Mech 454, **203–233**. https://doi.org/10.1017/S0022112001007078

[6]   Sun, J., Burns, S.P., Vandemark, D., Donelan, M.A., Mahrt, L., Crawford, T.L., Herbers, T.H.C., Crescenti, G.H., French, J.R., 2005. Measurement of Directional Wave Spectra Using Aircraft Laser Altimeters. J Atmos Ocean Technol 22, **869–885**. https://doi.org/10.1175/JTECH1729.1

[7]   Harald E, 2005. Measuring and analysing the directional spectra of ocean waves. (Luxembourg: Office for Official Publications of the European Communities) .

[8]   Bourdier, S., Dampney, K., Lopez, H.F.G., Richon, J.-B., 2014. Non-intrusive wave field measurement. MARINET Report.

[9]   Benetazzo, A., 2006. Measurements of short water waves using stereo matched image sequences. Coastal Engineering 53, **1013–1032**. https://doi.org/10.1016/j.coastaleng.2006.06.012

[10]   Benetazzo, A., Fedele, F., Gallego, G., Shih, P.-C., Yezzi, A., 2012. Offshore stereo measurements of gravity waves. Coastal Engineering 64, **127–138**. https://doi.org/10.1016/j.coastaleng.2012.01.007

[11]   Guimarães, P.V., Ardhuin, F., Bergamasco, F., Leckler, F., Filipot, J.-F., Shim, J.-S., Dulov, V., Benetazzo, A., 2020. A data set of sea surface stereo images to resolve space-time wave fields. Sci Data 7, **145**. https://doi.org/10.1038/s41597-020-0492-9

[12]   Benetazzo, A., Serafino, F., Bergamasco, F., Ludeno, G., Ardhuin, F., Sutherland, P., Sclavo, M., Barbariol, F., 2018. Stereo imaging and X-band radar wave data fusion: An assessment. Ocean Engineering 152, **346–352**. https://doi.org/10.1016/j.oceaneng.2018.01.077

[13]   Ginio N, Liberzon D, Lindenbaum M and Fishbain B 2024 Polarimetric images of mechanically generated water waves coupled with surface elevation measurements *ScienceDB. https://doi.org/10.57760/sciencedb.13968*

[14]   Hasselmann, K., 1973. Measurements of Wind-Wave Growth and Swell Decay during the Joint North Sea Wave Project (JONSWAP). Ergnzungsheft Zur Deutschen Hydrographischen Zeitschrift Reihe, A(8)(8 0), p.95. https://doi.org/citeulike-article-id:2710264 **46**.

[15]   Geiger, A., Moosmann, F., Car, O., Schuster, B., 2012. Automatic camera and range sensor calibration using a single shot, in: 2012 IEEE International Conference on Robotics and Automation. IEEE, pp. **3936–3943**. https://doi.org/10.1109/ICRA.2012.6224570

[16]   Solodoch, A., 2024. Retrieval of Surface Waves Spectrum from UAV Nadir Video. https://doi.org/10.36227/techrxiv.170854748.81779222/v1